%% file: 00-main.tex
\documentclass{article}

\usepackage{arxiv}
\usepackage[utf8]{inputenc}
\usepackage[T1]{fontenc}
\usepackage[inline]{enumitem}
\usepackage{booktabs}
\usepackage{amsmath}
\usepackage{amsfonts}
\usepackage{amssymb}
\usepackage{enumitem}
\usepackage{mathbbol}
\usepackage{mathtools}
\usepackage{hyperref}
\usepackage{url}
\usepackage{nicefrac}
\usepackage{microtype}
\usepackage{natbib}
\usepackage{multirow}
\usepackage{subcaption}

\newcommand{\ie}{\emph{i.e.}}
\newcommand{\eg}{\emph{e.g.}}

\title{Conformer-Kernel with Query Term Independence at TREC 2020 Deep Learning Track}

\author{
  Bhaskar Mitra \\
  Microsoft, University College London \\
  \texttt{bmitra@microsoft.com} \\
   \And
  Sebastian Hofst\"{a}tter \\
  TU Wien \\
  \texttt{s.hofstaetter@tuwien.ac.at}
   \And
  Hamed Zamani\thanks{Work done while at Microsoft.} \\
  University of Massachusetts Amherst \\
  \texttt{zamani@cs.umass.edu}
   \And
 Nick Craswell \\
  Microsoft \\
  \texttt{nickcr@microsoft.com} \\
}

\begin{document}
\maketitle

\input{01-abstract}


\input{02-intro}
\input{03-task}
\input{04-model}
\input{05-result}

\input{06-conclusion}

\bibliographystyle{plainnat}
\bibliography{bibtex}

\end{document}

%% file: 01-abstract.tex
\begin{abstract}
We benchmark Conformer-Kernel models under the strict blind evaluation setting of the TREC 2020 Deep Learning track.
In particular, we study the impact of incorporating:
\begin{enumerate*}[label=(\roman*)]
    \item Explicit term matching to complement matching based on learned representations (\ie, the ``Duet principle''),
    \item query term independence (\ie, the ``QTI assumption'') to scale the model to the full retrieval setting, and
    \item the ORCAS click data as an additional document description field.
\end{enumerate*}
We find evidence which supports that all three aforementioned strategies can lead to improved retrieval quality.
\end{abstract}

%% file: 02-intro.tex
\section{Introduction}
\label{sec:intro}

The Conformer-Kernel (CK) model~\citep{mitra2020conformer} builds upon the Transformer-Kernel (TK)~\citep{hofstatter2019tu} architecture, that demonstrated strong competitive performance compared to BERT-based~\citep{devlin2018bert} ranking methods, but notably at a fraction of the compute and GPU memory cost, at the TREC 2019 Deep Learning track~\citep{craswell2019overview}.
Notwithstanding these strong results, the TK model suffers from two clear deficiencies.
Firstly, because the TK model employs stacked Transformers for query and document encoding, it is challenging to incorporate long body text into this model as the GPU memory requirement of Transformers' self-attention layers grows quadratically with respect to input sequence length.
So, for example, to increase the limit on the maximum input sequence length by $4\times$ from $128$ to $512$ we would require $16\times$ more GPU memory for each of the self-attention layers in the model.
Considering that documents can contain thousands of terms, this limits the model to inspecting only a subset of the document text which may have negative implications, such as poorer retrieval quality and under-retrieval of longer documents~\citep{hofstatter2020improving}.
Secondly, the original TK model was designed for the reranking task and requires that every document in a given candidate set be evaluated individually with respect to the query.
This is problematic if we want to use the model to retrieve from the full collection which may contain millions, if not billions, of documents.
\citet{zamani2018neural} raised this concern for the first time and addressed it by learning sparse representations for query and documents for inverted indexing. Later, \citet{mitra2019incorporating} proposed an alternative solution based on the query term independence (QTI) assumption, which was adopted by \citet{mitra2020conformer}. They replaced the Transformer layers with novel Conformer counterparts and incorporated the QTI assumption into the model design.

In their original paper, \citet{mitra2020conformer} compared their model to other retrieval methods, under the full retrieval setting, based on the test set from the TREC 2019 Deep Learning track~\citep{craswell2019overview} for which both the queries and relevance labels are currently available publicly.
This evaluation is less stringent than participating in the official annual TREC benchmarking because:
\begin{enumerate*}[label=(\alph*)]
    \item it allows the experimenter to run multiple evaluations against the test set which may lead to overfitting, and
    \item it uses pre-collected labels which may not cover additional relevant documents that a new model may surface and consequently under-report the performance of dramatically new approaches~\citep{yilmaz2020reliability}.
\end{enumerate*}
Therefore, in this work, we evaluate the model under the stricter TREC benchmarking setting in the 2020 edition of the Deep Learning track~\citep{craswell2020overview}.

%% file: 03-task.tex
\section{TREC 2020 Deep Learning track}
\label{sec:task}

The TREC 2020 Deep Learning track~\citep{craswell2020overview} uses the same training data as the previous year~\citep{craswell2019overview}, which was originally derived from the MS MARCO dataset~\citep{bajaj2016ms}.
However, the track provides a new blind test set for the second year.
In our work, we only consider the document ranking task, although the track also allows participants to evaluate their models on passage ranking.
The training data for the document ranking task consists of $384,597$ positively labeled query-document pairs.
The test set comprised of $200$ queries out of which $45$ queries were selected by NIST for judging.
We report four relevance metrics---NDCG@10~\citep{JK2002}, NCG@100~\citep{rosset2018optimizing}, AP~\citep{zhu2004recall}, and RR~\citep{craswell2009mean}---computed over these $45$ queries.
Under the \emph{rerank} setting, each model is expected to re-order a set of $100$ candidate documents provided per query, and under the \emph{fullrank} setting each model must retrieve a ranked list of maximum hundred documents from a collection containing $3,213,835$ documents in response to each query.

%% file: 04-model.tex
\section{Conformer-Kernel with Query Term Independence}
\label{sec:model}

\begin{figure}
\center
\includegraphics[width=\textwidth]{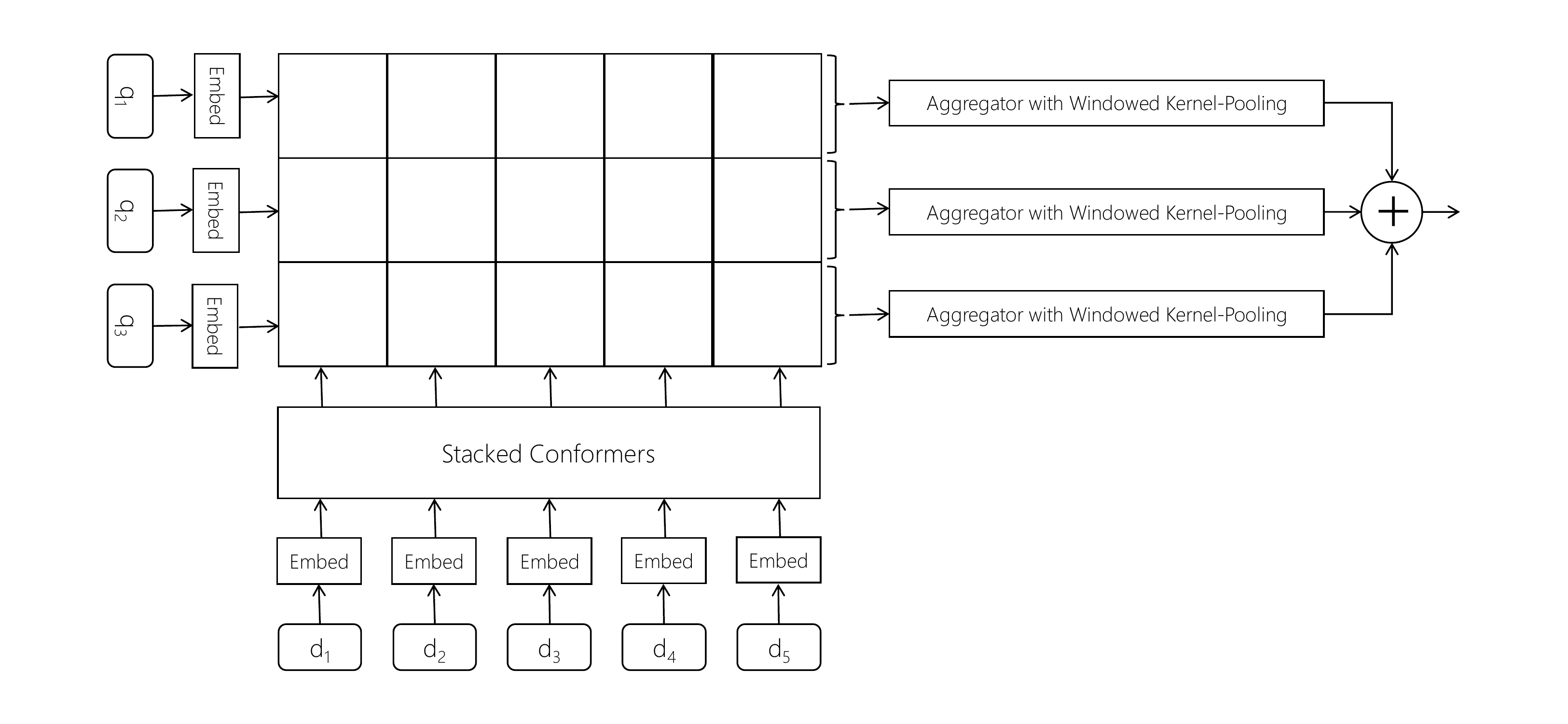}
\caption{The NDRM1 variant of the CK model with QTI.}
\label{fig:model}
\end{figure}

The CK models combine novel Conformer layers with several other existing ideas from the neural information retrieval literature~\citep{mitra2021neural, mitra2018introduction,guo2020deep}.
We use the publicly available implementation\footnote{\url{https://github.com/bmitra-msft/TREC-Deep-Learning-Quick-Start}} of CK models in our work, and adopt the same model taxonomy as in the code to describe the different variants.

The \emph{NDRM1} variant builds on the TK architecture~\citep{hofstatter2019tu} by incorporating two key changes:
\begin{enumerate*}[label=(\roman*)]
    \item It replaces the Transformer layers with Conformer layers, and
    \item factorizes the model to incorporate the QTI assumption.
\end{enumerate*}
Figure~\ref{fig:model} visualizes the NDRM1 architecture.
Unlike other attempts~\citep{hofstatter2020improving} at extending the TK architecture to long text by treating the document as a collection of passages, the Conformer layer replaces the standard self-attention mechanism with a separable self-attention mechanism whose memory complexity of $\mathcal{O}(n \times d_\text{key})$---where $n$ is input sequence length and $d_\text{key}$ is the size of the learned key embeddings---is a significant improvement over the quadratic $\mathcal{O}(n^2)$ complexity of standard self-attention.
Furthermore, the Conformer layer complements the self-attention with an additional convolutional layer to more accurately model local context within the text.
Next, to incorporate query term independence, the model evaluates the relevance of the document to each query term independently and then linearly combines those relevance estimates to obtain the aggregated estimate for the full query.
By incorporating the QTI assumption, we can precompute all term-document scores at indexing time and employ an inverted index data structure to perform fast retrieval at query time.

The \emph{NDRM2} model can be described as a learned relevance function that only inspects the count of exact matches of query terms in the document and bears a similar form as BM25~\citep{robertson2009probabilistic}.
Similar to BM25, the NDRM2 model is also compliant with the QTI assumption.
A linear combination of NDRM1 and NDRM2 gives us the NDRM3 model.
This strategy of combining an exact term matching subnetwork with a representation learning based matching subnetwork has been previously studied in the context of the Duet architecture~\citep{mitra2017learning, mitra2019duet, nanni2017benchmark, mitra2019updated}, and have been reported to be specifically effective under the full retrieval setting~\citep{mitra2016desm, mitra2020conformer, kuzi2020leveraging, gao2020complementing, wrzalik2020cort}.
Because of the limit on the number of run submission to TREC, we only evaluate the NDRM1 and NDRM3 models in this work, although we have confirmed on the TREC 2019 test set that the NDRM2 model is competitive with a well-tuned BM25 baseline.

For the second edition of the TREC Deep Learning track, participants were also provided a click log dataset called ORCAS~\citep{craswell2020orcas} that can be used in any way the participants deem appropriate.
We use clicked queries in the ORCAS data as additional meta description for the corresponding documents to complement the intrinsic document content in the form of URL, title, and body text.
While previous work~\citep{zamani2018neural2} have explored using fielded document input representation in the context of deep neural ranking models, in this work we simply concatenate the text from different fields to produce a flat unstructured input representation of the document that is fed into the model.

We test each model variant under both the rerank and the fullrank settings of the document ranking task in the Deep Learning track.
We use the same hyperparameters and other configuration settings as prescribed by \citet{mitra2020conformer}.

%% file: 05-result.tex
\section{Results}
\label{sec:result}

\begin{table}
    \small
    \centering
    \caption{Official TREC results.
    All metrics are computed at a rank threshold of 100, unless explicitly specified.}
    \begin{tabular}{lllllccccc}
    \hline
    \hline
        \textbf{Run description} & \textbf{Run ID} & \textbf{Subtask} & \textbf{NDCG@10} & \textbf{NCG@100} & \textbf{AP} & \textbf{RR} \\
        \hline
        NDRM1 & ndrm1-full & fullrank & $0.5991$  & $0.6280$ & $0.3858$ & $0.9333$ \\
        NDRM1 & ndrm1-re & rerank & $0.6161$ & $0.6283$ & $0.4150$ & $0.9333$ \\
        NDRM3 & ndrm3-re & rerank & $0.6162$ & $0.6283$ & $0.4122$ & $0.9333$ \\
        NDRM3 & ndrm3-full & fullrank & $0.6162$ & $0.6626$ & $0.4069$ & $0.9333$ \\
        NDRM3 + ORCAS & ndrm3-orc-re & rerank & $0.6217$ & $0.6283$ & $0.4194$ & $0.9241$ \\
        NDRM3 + ORCAS & ndrm3-orc-full & fullrank & $0.6249$ & $0.6764$ & $0.4280$ & $0.9444$ \\
        \hline
        \hline
    \end{tabular}
    \label{tbl:results}
\end{table}

\begin{figure}
  \center
  \begin{subfigure}{.9\textwidth}
    \includegraphics[width=\textwidth]{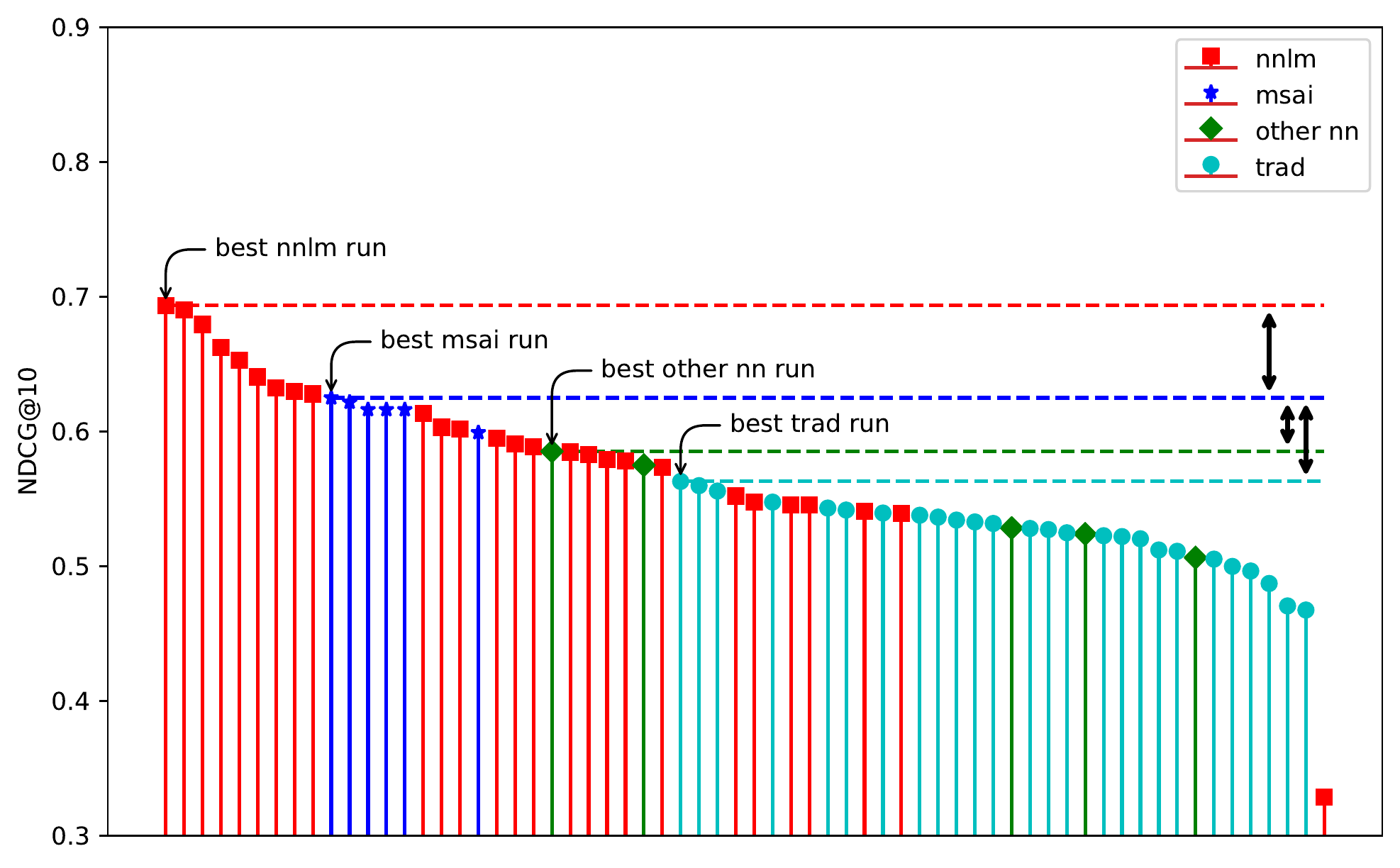}
    \caption{NDCG@10}
    \label{fig:competitive-ndcg}
  \end{subfigure}
  \hfill
  \begin{subfigure}{.9\textwidth}
    \includegraphics[width=\textwidth]{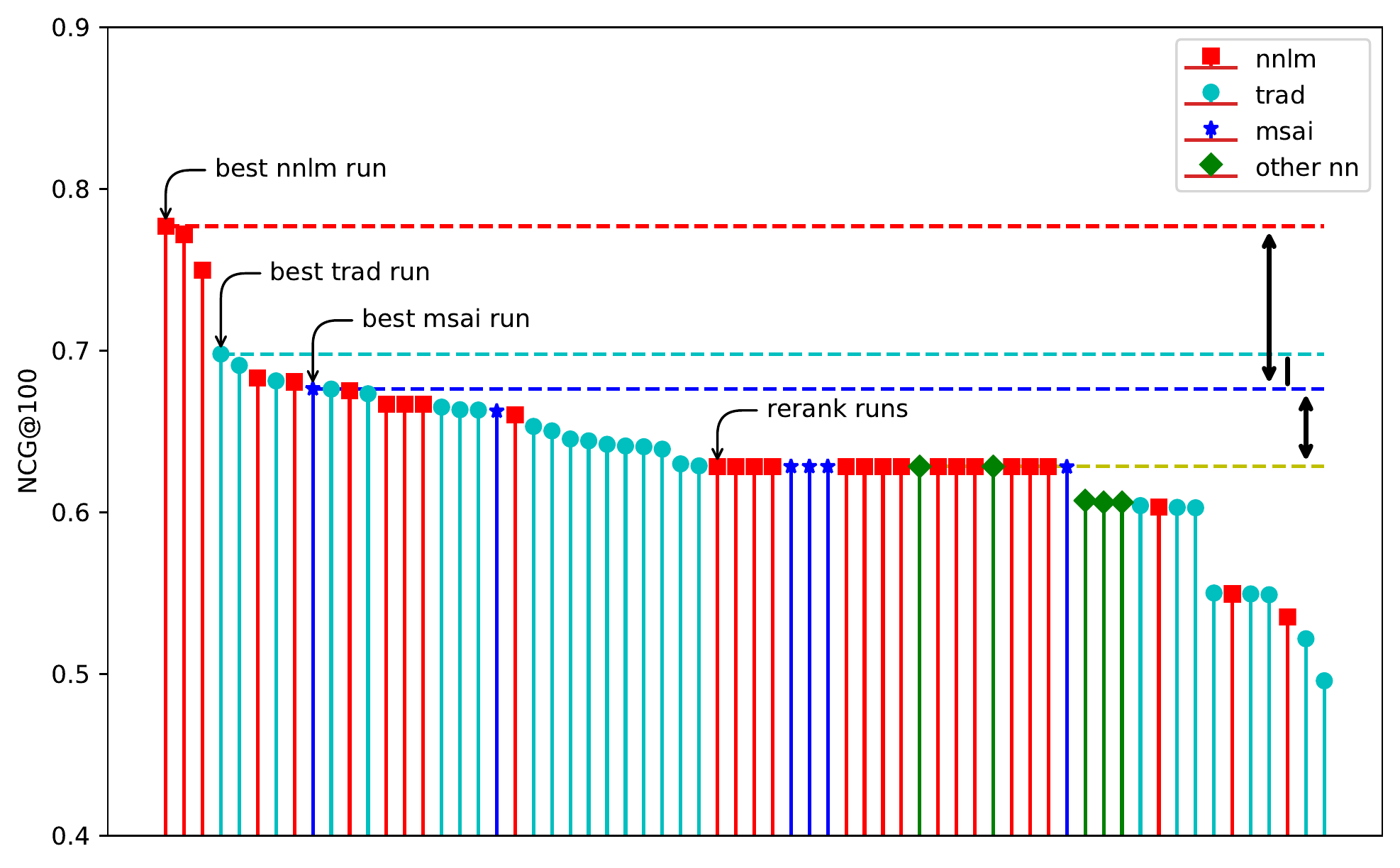}
    \caption{NCG@100}
    \label{fig:competitive-ncg}
  \end{subfigure}
  \caption{Comparing our runs with runs submitted by other groups. We adopt the same ``nnlm'', ``nn'', and ``trad'' taxonomy for models as in the track overview~\citep{craswell2020overview}. All our runs are ``nn'' runs under this classification but we label them specifically as ``msai'' to distinguish from ``other nn runs''. The runs in each plot are sorted independently based on the corresponding metric.}
  \label{fig:competitive}
\end{figure}

\begin{figure}
\includegraphics[width=.95\textwidth]{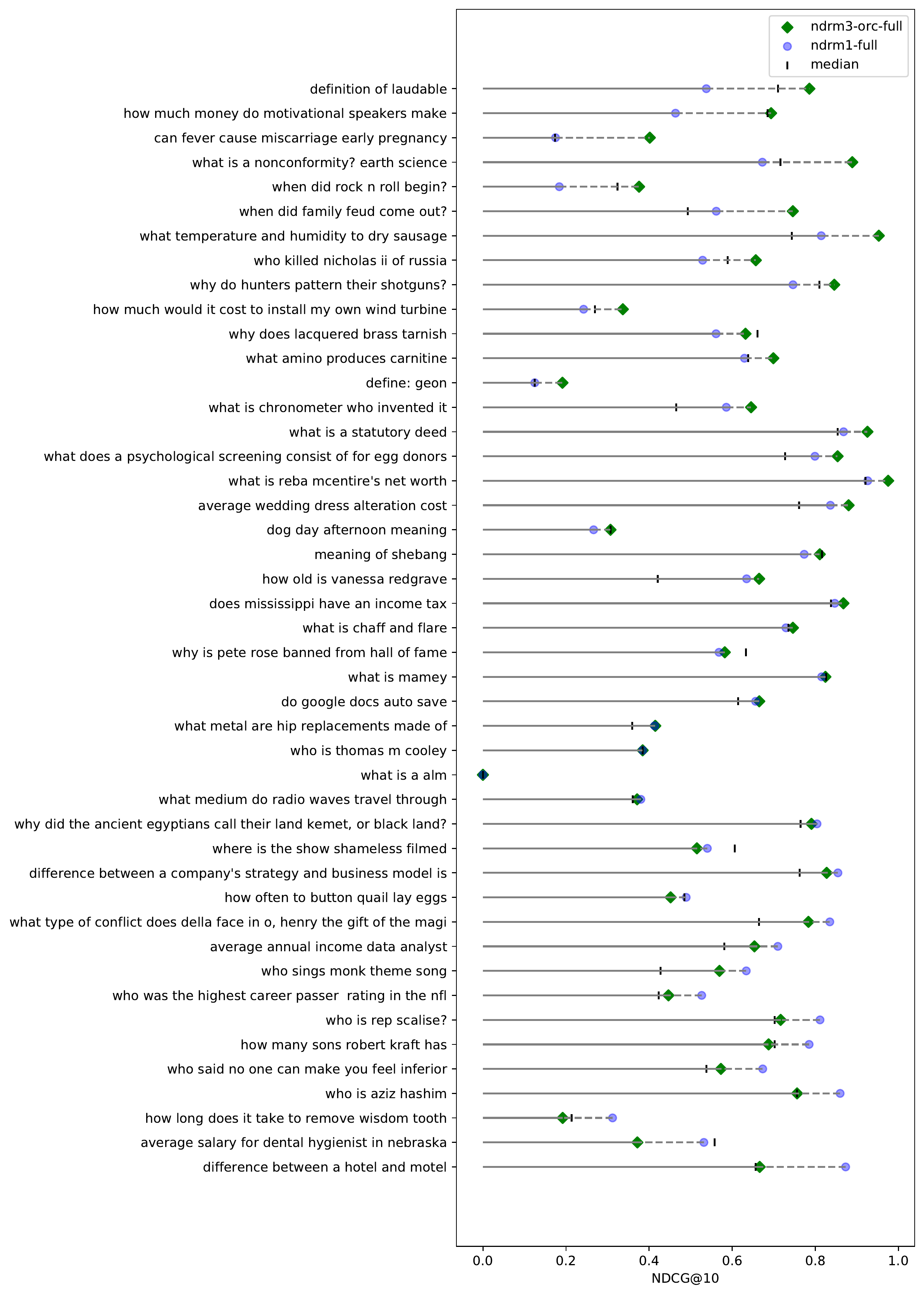}
\caption{Per-query comparison between our worst performing run (``ndrm1-full'') and our best performing run (``ndrm3-orc-full'') based on the NDCG@10 metric.
Median NDCG@10 across all track submissions also shown for reference.}
\label{fig:ndrm3-orc-full-v-ndrm1-full-per-query}
\end{figure}

\begin{figure}
\includegraphics[width=\textwidth]{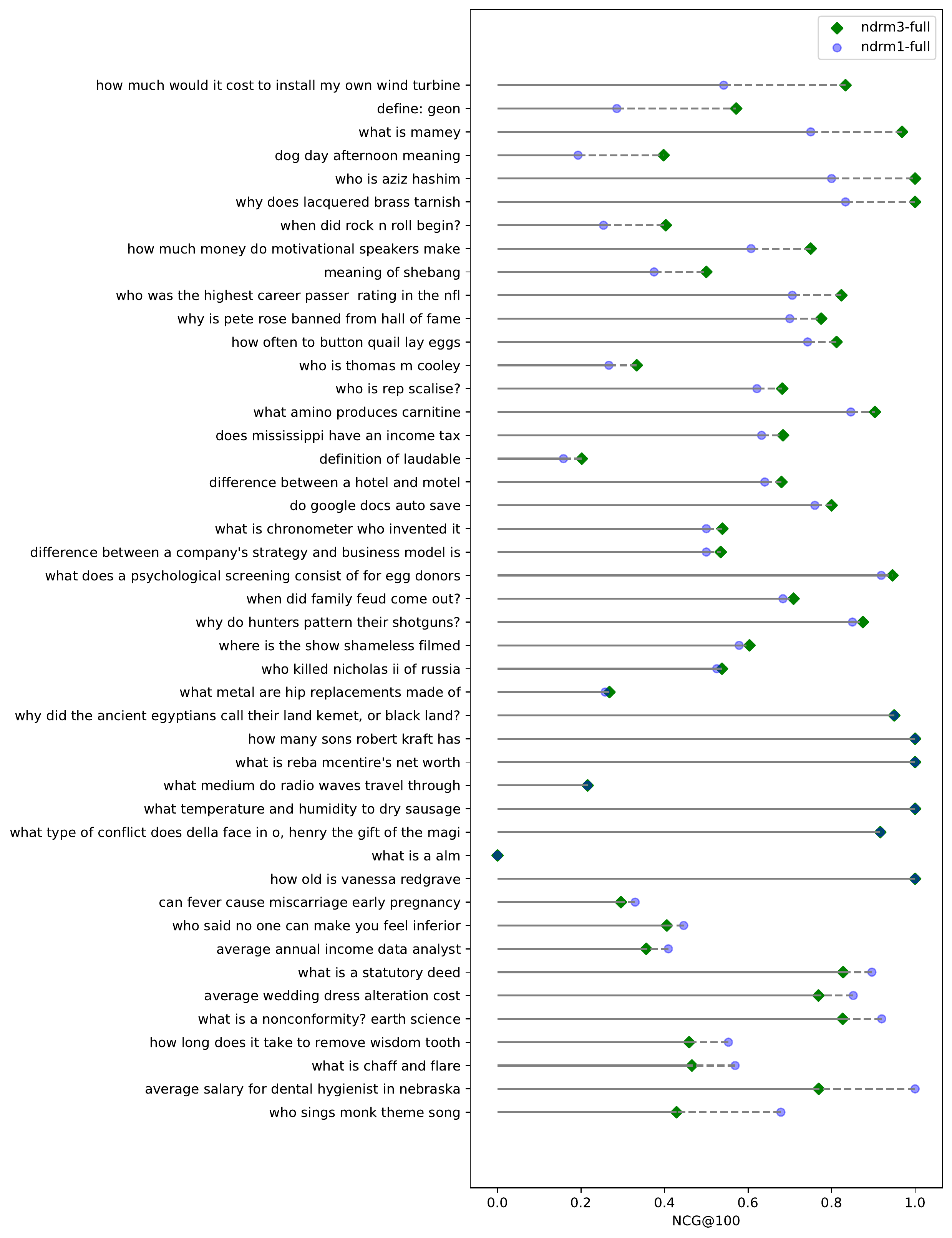}
\caption{Per-query comparison between the ``ndrm1-full'' and the ``ndrm3-full'' runs based on the NCG@100 metric.}
\label{fig:ndrm3-v-ndrm1-per-query}
\end{figure}

This year at TREC, we focus our study on the following four research questions.

\paragraph{RQ1. How does our best run perform competitively compared to other submissions?}

Table~\ref{tbl:results} summarizes the relevance metrics corresponding to all the submitted runs.
According to the taxonomy proposed by \citet{craswell2019overview}, the CK models can be described as ``nn'' models---\ie, neural models without large scale pretraining as has been popularized by models like BERT~\citep{devlin2018bert}.
Figure~\ref{fig:competitive} shows that our best run ``ndrm3-orc-full'' was also the best performing ``nn'' run on both NDCG@10 and NCG@100.
Furthermore, on NDCG@10 our best run outperforms two-third of the ``nnlm'' runs while also requiring significantly less resources to train and evaluate compared to those models.
It is also noteworthy, that ``ndrm3-orc-full'' employs a single-stage ranking, whereas all the runs that outperform it implement some form of cascaded ranking~\citep{wang2011cascade, matveeva2006high} with multiple rank-and-prune stages.
With respect to the full retrieval setting, we note that ``ndrm3-orc-full'' improves NCG@100 by $+0.0481$ over the provided candidates for the reranking setting, which puts it among the 10 top performing runs according to NCG@100.
Finally, Figure~\ref{fig:ndrm3-orc-full-v-ndrm1-full-per-query} shows the per-query performance of our best and worst performing runs compared to the median performance.
This figure provides further evidence that the CK models achieve competitive retrieval quality among all the track submissions this year.

\paragraph{RQ2. Does explicit term matching improve retrieval quality?}
To shed light on this question, we compare the NDRM1 and the NDRM3 models, where the only difference between the two models is that the latter incorporates the explicit term matching signal while former does not.
We find that under the reranking setting---\ie, when comparing the ``ndrm1-re'' and the ``ndrm3-re'' runs---there is no clear evidence that the explicit term matching is beneficial.
This is likely because the candidate documents for reranking were generated by a first-stage BM25 ranker and hence the explicit term matching signal is already part of the end-to-end retrieval stack.
However, under the fullrank setting---\ie, when comparing the ``ndrm1-full'' and the ``ndrm3-full'' runs---we see moderate improvements across all metrics: $2.9\%$ improvement in NDCG@10 and $5.5\%$ improvement in both AP and NCG@100.
These observations are supported by \citet{kuzi2020leveraging}, who find that exact term matching are important for the fullrank setting, and also by \citet{xiong2020approximate} who observe that their proposed model which does not incorporate exact matching fare better in the rerank setting than on the fullrank subtask.

Figure~\ref{fig:ndrm3-v-ndrm1-per-query} compares how the ``ndrm1-full'' and the ``ndrm3-full'' runs perform on the $45$ different queries in the test set.
Based on a qualitative inspection of the queries, it appears that exact term matching may be important for queries containing named entities---\eg, ``who is \emph{aziz hashim}'' and ``why is \emph{pete rose} banned from hall of fame''---where it is necessary to ensure that the retrieved documents are about the correct entity.

\paragraph{RQ3. How does the retrieval quality differ for our model between the fullrank and the rerank setting?}
As expected, we find that without exact term matching, the retrieval quality for CK models are lower under the fullrank setting compared to the rerank setting---\ie, ``ndrm1-re'' is better than ``ndrm1-full''.
In contrast, when exact term matching is incorporated, the CK model achieves $5.5\%$ improvement in NCG, which is a recall-oriented metric, in the fullrank setting (``ndrm3-full'') compared to its counterpart under the rerank setting (``ndrm3-re'').
However, on all the other metrics we see no difference (NDCG@10 and RR) or small regression ($1.3\%$ for AP) under the fullrank setting.
Finally, if we introduce the ORCAS data---\ie, compare ``ndrm3-orc-full'' and ``ndrm3-orc-re''---we see improvements under the fullrank setting across all metrics: $7.7\%$ for NCG@100, $2.2\%$ for RR, $2.1\%$ for AP, and $0.5\%$ for NDCG@10.

In adhoc retrieval, a common strategy involves sequentially cascading multiple rank-and-prune stages~\citep{matveeva2006high, wang2011cascade, chen2017efficient, gallagher2019joint, nogueira2019multi} for better effectiveness-efficiency trade-offs.
Following a similar strategy, we may be able to improve on these results by introducing additional reranking stages on top of a first stage retrieval using query term independent CK models.
We anticipate that this may be an interesting area for future exploration.

\paragraph{RQ4. Does using ORCAS queries as an additional document description field improve retrieval quality?}
Finally, we want to study if the incorporation of click log datasets, such as ORCAS~\citep{craswell2020orcas}, can be beneficial for retrieval quality.
We find that on the rerank subtask, both NDCG@10 and AP improve by $0.9\%$ and $1.7\%$, respectively, although RR degrades by $1\%$.
On the fullrank subtask, the addition of ORCAS signal seems to improve all metrics: AP by $5.2\%$, NCG@100 by $2.1\%$, NDCG@10 by $1.4\%$, and RR by $1.2\%$.
These results indicate that ORCAS, and other similar click log datasets, may be useful for achieving better retrieval relevance.

%% file: 06-conclusion.tex
\section{Conclusion}
\label{sec:conclusion}
In this work, we benchmark CK models under the strict blind evaluation setting of the TREC 2020 Deep Learning track.
We find that our best CK model outperforms all ``trad'' runs, remaining ``nn'' runs, and two-thirds of ``nnlm'' runs this year.
In addition to strong competitive performance, we find that incorporating
\begin{enumerate*}[label=(\roman*)]
    \item exact term matching (the ``Duet principle''),
    \item query term independence (the ``QTI assumption''), and
    \item ORCAS data as an additional document field
\end{enumerate*}
all generally contribute positively to retrieval quality, and the run ``ndrm3-orc-full'' that incorporates all three techniques achieves our best performance.
We posit that considering the significantly lower cost of training and evaluating CK models, these models provide interesting alternatives to BERT-based rankers with different operating points on the effectiveness-efficiency curve.

%% file: 00-main.bbl
\begin{thebibliography}{34}
\providecommand{\natexlab}[1]{#1}
\providecommand{\url}[1]{\texttt{#1}}
\expandafter\ifx\csname urlstyle\endcsname\relax
  \providecommand{\doi}[1]{doi: #1}\else
  \providecommand{\doi}{doi: \begingroup \urlstyle{rm}\Url}\fi

\bibitem[Bajaj et~al.(2016)Bajaj, Campos, Craswell, Deng, Gao, Liu, Majumder,
  McNamara, Mitra, Nguyen, et~al.]{bajaj2016ms}
Payal Bajaj, Daniel Campos, Nick Craswell, Li~Deng, Jianfeng Gao, Xiaodong Liu,
  Rangan Majumder, Andrew McNamara, Bhaskar Mitra, Tri Nguyen, et~al.
\newblock {MS MARCO: A Human Generated MAchine Reading COmprehension Dataset}.
\newblock \emph{arXiv preprint arXiv:1611.09268}, 2016.

\bibitem[Chen et~al.(2017)Chen, Gallagher, Blanco, and
  Culpepper]{chen2017efficient}
Ruey-Cheng Chen, Luke Gallagher, Roi Blanco, and J~Shane Culpepper.
\newblock {Efficient Cost-Aware Cascade Ranking in Multi-Stage Retrieval}.
\newblock In \emph{Proc. of SIGIR}, 2017.

\bibitem[Craswell(2009)]{craswell2009mean}
Nick Craswell.
\newblock {Mean Reciprocal Rank}.
\newblock In \emph{Encyclopedia of Database Systems}, pages 1703--1703.
  Springer, 2009.

\bibitem[Craswell et~al.(2020{\natexlab{a}})Craswell, Campos, Mitra, Yilmaz,
  and Billerbeck]{craswell2020orcas}
Nick Craswell, Daniel Campos, Bhaskar Mitra, Emine Yilmaz, and Bodo Billerbeck.
\newblock {ORCAS: 18 Million Clicked Query-Document Pairs for Analyzing
  Search}.
\newblock In \emph{Proc. of CIKM}, 2020{\natexlab{a}}.

\bibitem[Craswell et~al.(2020{\natexlab{b}})Craswell, Mitra, Yilmaz, and
  Campos]{craswell2019overview}
Nick Craswell, Bhaskar Mitra, Emine Yilmaz, and Daniel Campos.
\newblock {Overview of the TREC 2019 Deep Learning Track}.
\newblock In \emph{Proc. of TREC}, 2020{\natexlab{b}}.

\bibitem[Craswell et~al.(2020{\natexlab{c}})Craswell, Mitra, Yilmaz, and
  Campos]{craswell2020overview}
Nick Craswell, Bhaskar Mitra, Emine Yilmaz, and Daniel Campos.
\newblock {Overview of the TREC 2020 Deep Learning Track}.
\newblock In \emph{Proc. of TREC (to be published)}, 2020{\natexlab{c}}.

\bibitem[Devlin et~al.(2019)Devlin, Chang, Lee, and Toutanova]{devlin2018bert}
Jacob Devlin, Ming-Wei Chang, Kenton Lee, and Kristina Toutanova.
\newblock {BERT: Pre-training of Deep Bidirectional Transformers for Language
  Understanding}.
\newblock In \emph{Proc. of NAACL}, 2019.

\bibitem[Gallagher et~al.(2019)Gallagher, Chen, Blanco, and
  Culpepper]{gallagher2019joint}
Luke Gallagher, Ruey-Cheng Chen, Roi Blanco, and J~Shane Culpepper.
\newblock {Joint Optimization of Cascade Ranking Models}.
\newblock In \emph{Proc. of WSDM}, 2019.

\bibitem[Gao et~al.(2020)Gao, Dai, Fan, and Callan]{gao2020complementing}
Luyu Gao, Zhuyun Dai, Zhen Fan, and Jamie Callan.
\newblock {Complementing Lexical Retrieval with Semantic Residual Embedding}.
\newblock \emph{arXiv preprint arXiv:2004.13969}, 2020.

\bibitem[Guo et~al.(2020)Guo, Fan, Pang, Yang, Ai, Zamani, Wu, Croft, and
  Cheng]{guo2020deep}
Jiafeng Guo, Yixing Fan, Liang Pang, Liu Yang, Qingyao Ai, Hamed Zamani, Chen
  Wu, W.~Bruce Croft, and Xueqi Cheng.
\newblock {A Deep Look into Neural Ranking Models for Information Retrieval}.
\newblock \emph{IP\&M}, 2020.

\bibitem[Hofst{\"a}tter et~al.(2019)Hofst{\"a}tter, Zlabinger, and
  Hanbury]{hofstatter2019tu}
Sebastian Hofst{\"a}tter, Markus Zlabinger, and Allan Hanbury.
\newblock {TU Wien@ TREC Deep Learning'19--Simple Contextualization for
  Re-ranking}.
\newblock In \emph{Proc. of TREC}, 2019.

\bibitem[Hofst{\"a}tter et~al.(2020)Hofst{\"a}tter, Zamani, Mitra, Craswell,
  and Hanbury]{hofstatter2020improving}
Sebastian Hofst{\"a}tter, Hamed Zamani, Bhaskar Mitra, Nick Craswell, and Allan
  Hanbury.
\newblock {Local Self-Attention over Long Text for Efficient Document
  Retrieval}.
\newblock In \emph{Proc. of SIGIR}. ACM, 2020.

\bibitem[J\"arvelin and Kek\"al\"ainen(2002)]{JK2002}
K.~J\"arvelin and J.~Kek\"al\"ainen.
\newblock Cumulated gain-based evaluation of {IR} techniques.
\newblock \emph{ACM TOIS}, 20\penalty0 (4):\penalty0 422--446, 2002.

\bibitem[Kuzi et~al.(2020)Kuzi, Zhang, Li, Bendersky, and
  Najork]{kuzi2020leveraging}
Saar Kuzi, Mingyang Zhang, Cheng Li, Michael Bendersky, and Marc Najork.
\newblock {Leveraging Semantic and Lexical Matching to Improve the Recall of
  Document Retrieval Systems: A Hybrid Approach}.
\newblock \emph{arXiv preprint arXiv:2010.01195}, 2020.

\bibitem[Matveeva et~al.(2006)Matveeva, Burges, Burkard, Laucius, and
  Wong]{matveeva2006high}
Irina Matveeva, Chris Burges, Timo Burkard, Andy Laucius, and Leon Wong.
\newblock {High Accuracy Retrieval with Multiple Nested Ranker}.
\newblock In \emph{Proc. of SIGIR}, 2006.

\bibitem[Mitra(2021)]{mitra2021neural}
Bhaskar Mitra.
\newblock \emph{Neural Methods for Effective, Efficient, and Exposure-Aware
  Information Retrieval}.
\newblock PhD thesis, University College London, 2021.

\bibitem[Mitra and Craswell(2018)]{mitra2018introduction}
Bhaskar Mitra and Nick Craswell.
\newblock {An Introduction to Neural Information Retrieval}.
\newblock \emph{Foundations and Trends{\textregistered} in Information
  Retrieval}, 2018.

\bibitem[Mitra and Craswell(2019{\natexlab{a}})]{mitra2019duet}
Bhaskar Mitra and Nick Craswell.
\newblock {Duet at TREC 2019 Deep Learning Track}.
\newblock 2019{\natexlab{a}}.

\bibitem[Mitra and Craswell(2019{\natexlab{b}})]{mitra2019updated}
Bhaskar Mitra and Nick Craswell.
\newblock {An Updated Duet Model for Passage Re-ranking}.
\newblock \emph{arXiv preprint arXiv:1903.07666}, 2019{\natexlab{b}}.

\bibitem[Mitra et~al.(2016)Mitra, Nalisnick, Craswell, and
  Caruana]{mitra2016desm}
Bhaskar Mitra, Eric Nalisnick, Nick Craswell, and Rich Caruana.
\newblock {A Dual Embedding Space Model for Document Ranking}.
\newblock \emph{arXiv preprint arXiv:1602.01137}, 2016.

\bibitem[Mitra et~al.(2017)Mitra, Diaz, and Craswell]{mitra2017learning}
Bhaskar Mitra, Fernando Diaz, and Nick Craswell.
\newblock {Learning to Match Using Local and Distributed Representations of
  Text for Web Search}.
\newblock In \emph{Proc. of WWW}, pages 1291--1299, 2017.

\bibitem[Mitra et~al.(2019)Mitra, Rosset, Hawking, Craswell, Diaz, and
  Yilmaz]{mitra2019incorporating}
Bhaskar Mitra, Corby Rosset, David Hawking, Nick Craswell, Fernando Diaz, and
  Emine Yilmaz.
\newblock {Incorporating Query Term Independence Assumption for Efficient
  Retrieval and Ranking using Deep Neural Networks}.
\newblock \emph{arXiv preprint arXiv:1907.03693}, 2019.

\bibitem[Mitra et~al.(2020)Mitra, Hofstatter, Zamani, and
  Craswell]{mitra2020conformer}
Bhaskar Mitra, Sebastian Hofstatter, Hamed Zamani, and Nick Craswell.
\newblock {Conformer-Kernel with Query Term Independence for Document
  Retrieval}.
\newblock \emph{arXiv preprint arXiv:2007.10434}, 2020.

\bibitem[Nanni et~al.(2017)Nanni, Mitra, Magnusson, and
  Dietz]{nanni2017benchmark}
Federico Nanni, Bhaskar Mitra, Matt Magnusson, and Laura Dietz.
\newblock {Benchmark for Complex Answer Retrieval}.
\newblock In \emph{Proc. ICTIR}. ACM, 2017.

\bibitem[Nogueira et~al.(2019)Nogueira, Yang, Cho, and Lin]{nogueira2019multi}
Rodrigo Nogueira, Wei Yang, Kyunghyun Cho, and Jimmy Lin.
\newblock {Multi-Stage Document Ranking with BERT}.
\newblock \emph{arXiv preprint arXiv:1910.14424}, 2019.

\bibitem[Robertson et~al.(2009)Robertson, Zaragoza,
  et~al.]{robertson2009probabilistic}
Stephen Robertson, Hugo Zaragoza, et~al.
\newblock {The Probabilistic Relevance Framework: BM25 and Beyond}.
\newblock \emph{Foundations and Trends{\textregistered} in Information
  Retrieval}, 3\penalty0 (4):\penalty0 333--389, 2009.

\bibitem[Rosset et~al.(2018)Rosset, Jose, Ghosh, Mitra, and
  Tiwary]{rosset2018optimizing}
Corby Rosset, Damien Jose, Gargi Ghosh, Bhaskar Mitra, and Saurabh Tiwary.
\newblock {Optimizing Query Evaluations using Reinforcement Learning for Web
  Search}.
\newblock In \emph{Proc. of SIGIR}, 2018.

\bibitem[Wang et~al.(2011)Wang, Lin, and Metzler]{wang2011cascade}
Lidan Wang, Jimmy Lin, and Donald Metzler.
\newblock {A Cascade Ranking Model for Efficient Ranked Retrieval}.
\newblock In \emph{Proc. of SIGIR}, 2011.

\bibitem[Wrzalik and Krechel(2020)]{wrzalik2020cort}
Marco Wrzalik and Dirk Krechel.
\newblock {CoRT: Complementary Rankings from Transformers}.
\newblock \emph{arXiv preprint arXiv:2010.10252}, 2020.

\bibitem[Xiong et~al.(2020)Xiong, Xiong, Li, Tang, Liu, Bennett, Ahmed, and
  Overwijk]{xiong2020approximate}
Lee Xiong, Chenyan Xiong, Ye~Li, Kwok-Fung Tang, Jialin Liu, Paul Bennett,
  Junaid Ahmed, and Arnold Overwijk.
\newblock {Approximate Nearest Neighbor Negative Contrastive Learning for Dense
  Text Retrieval}.
\newblock \emph{arXiv preprint arXiv:2007.00808}, 2020.

\bibitem[Yilmaz et~al.(2020)Yilmaz, Craswell, Mitra, and
  Campos]{yilmaz2020reliability}
Emine Yilmaz, Nick Craswell, Bhaskar Mitra, and Daniel Campos.
\newblock {On the Reliability of Test Collections for Evaluating Systems of
  Different Types}.
\newblock In \emph{Proc. of SIGIR}, 2020.

\bibitem[Zamani et~al.(2018{\natexlab{a}})Zamani, Dehghani, Croft,
  Learned-Miller, and Kamps]{zamani2018neural}
Hamed Zamani, Mostafa Dehghani, W~Bruce Croft, Erik Learned-Miller, and Jaap
  Kamps.
\newblock {From Neural Re-Ranking to Neural Ranking: Learning a Sparse
  Representation for Inverted Indexing}.
\newblock In \emph{Proc. of CIKM}, 2018{\natexlab{a}}.

\bibitem[Zamani et~al.(2018{\natexlab{b}})Zamani, Mitra, Song, Craswell, and
  Tiwary]{zamani2018neural2}
Hamed Zamani, Bhaskar Mitra, Xia Song, Nick Craswell, and Saurabh Tiwary.
\newblock {Neural Ranking Models with Multiple Document Fields}.
\newblock In \emph{Proc. of WSDM}, 2018{\natexlab{b}}.

\bibitem[Zhu(2004)]{zhu2004recall}
Mu~Zhu.
\newblock {Recall, Precision and Average Precision}.
\newblock \emph{Department of Statistics and Actuarial Science, University of
  Waterloo, Waterloo}, 2:\penalty0 30, 2004.

\end{thebibliography}
